\documentstyle[aaspp4,11pt]{article}
\received{1998 May 21}
\eqsecnum
\newcommand{\etal}{et~al.\ }
\newcommand{\eg}{e.g.,\ }

\slugcomment{Submitted to ApJ Letters}

\begin{document}

\title{SECULAR EVOLUTION OF BARRED GALAXIES WITH\\
MASSIVE CENTRAL BLACK HOLES}

\author{SHUNSUKE HOZUMI}
\affil{Faculty of Education, Shiga University, 2-5-1 Hiratsu,
       Otsu, Shiga 520-0862, Japan; hozumi@sue.shiga-u.ac.jp}

\author{LARS HERNQUIST}
\affil{Lick Observatory, University of California, Santa Cruz,
Santa Cruz, CA 95064; lars@ucolick.org}

\begin{abstract} 

The influence of central black holes on the dynamical evolution of
bars in disk galaxies is examined.  In particular, we use numerical
simulation to estimate the minimum mass black hole (BH) needed to
destroy a bar.  Initially, bars form in the disks via dynamical
instability.  Thereafter, once a bar is fully developed, a BH is
adiabatically added at the center of the disk.  To mitigate the
global effects of gravitational softening, Poisson's equation for 
the disk is solved by expanding the density and potential of the
galaxy in a set of basis functions.  

Our results indicate that a bar can be completely destroyed in a time
much smaller than a Hubble time if the central mass exceeds about 0.5\%
of the disk mass.  Since the implied minimum BH mass for bar destruction
is of order $10^{8.5}M_\odot$ for a typical disk galaxy, this process
should not be a rare phenomenon.  The bar amplitude decreases gradually
with time after the BH is added, and the rate at which the bar is
destroyed increases with increasing BH mass.  This suggests that bar
destruction arises from scattering of stars that support the bar as they
pass close to the center.

\end{abstract}

\keywords{celestial mechanics, stellar dynamics --- galaxies: structure
--- methods: numerical}

\vfil\eject

\section{INTRODUCTION}

Recent observations indicate that massive central black holes exist in
disk galaxies as well as in ellipticals.  For example, NGC 4945 (type
Sc), the Milky Way (type Sbc), NGC 1068 (type Sb), M31 (type Sb), NGC
4258 (type Sbc), and NGC 4594 (type Sa) are thought to harbor central
black holes with masses $\sim 10^6 M_\odot$, $2 \times 10^6 M_\odot$,
$10^7 M_\odot$, $3 \times 10^7 M_\odot$, $4 \times 10^7 M_\odot$, and
$5\times 10^8 M_\odot$, respectively (see, \eg Kormendy \&
Richstone 1995; van der Marel 1998; and references therein).  When
combined with data on nearby ellipticals (\eg Magorrian \etal
1998) it is plausible that most, if not all, large galaxies
contain central black holes whose masses range from $\sim 10^6M_\odot$
to $\sim 10^{9.5}M_\odot$.

Such a large mass concentration at the center of a galaxy could
affect the structure of the entire system.  Of great interest is the
influence of a central black hole (BH) on the structure of a bar, in
view of the fact that roughly half of all disk galaxies are barred.
Some studies (\eg Hasan \& Norman 1990; Hasan, Pfenniger, \& Norman
1993; Norman, Sellwood, \& Hasan 1996) have shown that central mass
concentrations can destroy a bar within a relatively short period of
time.  In particular, Norman \etal (1996) have concluded that a
central massive object with about 5\% the total mass of a disk plus
bulge can result in the dissolution of a bar within a Hubble time.
Remarkable as their result is, the implied mass for bar destruction
becomes about $10^{9.5}M_\odot$ when scaled to a typical disk galaxy
with a mass $\sim 10^{10.5} - 10^{11}M_\odot$.  If this mass
concentration is associated with a central black hole, the required BH
mass is greater than that inferred in nearby spirals and is comparable
to the largest BH masses derived observationally in ellipticals.  This
suggests that bar destruction by {\it central black holes} might be a
rare phenomenon, although this process could alternatively be driven by,
\eg sufficiently dense concentrations of gas.

In this paper, we employ N-body simulations to examine the influence
of a central black hole on a bar to determine how massive a BH is
required to destroy a bar.  To avoid the complications arising from
softening of gravitational forces, we employ a self-consistent field
(SCF) method, as termed by Hernquist \& Ostriker (1992), which solves
Poisson's equation by expanding the density and potential in a set of
basis functions.  Our results demonstrate, in fact, that in at least
some cases the minimum black hole mass for bar destruction may be at
least a factor of ten {\it smaller} than that suggested by the work of
Norman \etal (1996).  Of course, this is not necessarily in conflict
with the conclusions of Norman et~al., since they modeled the
consequences of the build-up of gas concentrations in the inner
regions of a barred disk, while we are specifically interested in the
influence of a central black hole.

\section{MODELS AND METHOD}

In the calculations described here, we study the evolution of
razor-thin exponential disks without bulges and halos, whose surface
density distributions are given by
\begin{equation} 
\mu(R)=\mu_0{\rm exp}(-R/h), 
\end{equation} 
where $h$ is the exponential scale-length and $R$ is the distance from
the center of the disk.  The disks are truncated at $R=15h$.  The full
phase-space is realized by employing the approach of Hernquist (1993),
who approximated the velocity distribution using moments of the
collisionless Boltzmann equation.  We choose parameters such that
the typical Toomre (1964) $Q$ parameter is of order unity, and the
models are globally unstable to the formation of bars.

For simplicity, the black holes are handled as external fields
and their potentials are approximated using a Plummer model given by
\begin{equation}
\phi_{\rm BH}(R)=-GM_\bullet(t)/\sqrt{R^2+\epsilon^2},
\end{equation}
where $G$, $M_\bullet(t)$, and $\epsilon$ are the gravitational
constant, BH mass, and scale-length of the potential, respectively.
The BH is added at $t=t_{\rm BH}$ long after the bar instability 
has occurred, and grows slowly from 0 to $M_{\rm BH}$ as follows:
\begin{equation}
M_\bullet (t)=\left\{
  \begin{array}{ll}
    M_{\rm BH}\left\{3\left[(t-t_{\rm BH})/t_{\rm grow}\right]^2
    -2\left[(t-t_{\rm BH})/t_{\rm grow}\right]^3\right\} &
    \quad {\rm for}\quad t_{\rm BH} \le t \le t_{\rm BH}+t_{\rm grow},\\
    M_{\rm BH} & \quad {\rm for}\quad t > t_{\rm BH}+t_{\rm grow},
  \end{array}\right.  \label{bhmass} 
\end{equation} 
where $t_{\rm grow}$ is the time for the BH to grow to its full
amplitude $M_{\rm BH}$.  Thus, the BH is made to grow adiabatically by
taking $t_{\rm grow}$ to be sufficiently long.  Here, we consider
cases in which $\epsilon=0.01 h$, and $M_{\rm BH}=0.01M$, $0.005M$,
and $0.001M$, where $M$ is the total mass of the disk.

In most of the experiments described below, we took $t_{\rm grow}=10$.
Identical calculations but with $t_{\rm grow}=5$ and 20 for $M_{\rm
BH}=0.01M$ yielded no practical differences in the subsequent evolution
from the choice $t_{\rm grow}=10$.  These values for $t_{\rm grow}$
can be compared with the typical rotation periods of the bars in the
simulations, $T_{\rm b}$.  To estimate $T_{\rm b}$, the phase angle
$\phi_{\rm b}(t)$ of the bar pattern is obtained from the phase of the
expansion coefficients $A_{22}(t)$ divided by 2 (see below).  Thus, the
bar rotation period can be calculated from the time derivative of
$\phi_{\rm b}(t)$.  We obtain $\Omega_{\rm b}=0.392$ between $t=60$ and
$t=100$ when there was no BH.  This means that the bar rotation period
is $T_{\rm b} = 2\pi/\Omega_{\rm b}=16.0$.  While our adopted values
for $t_{\rm grow}$ are not large compared to $T_{\rm b}$, the black
hole growth is adiabatic in the sense that the BH is added on a
timescale long compared with the dynamical times of stars near the
center of the disk.

Once the disks have been realized with particles, we evolve them
forward in time using an SCF method with Aoki \& Iye's (1978) basis
set, which is appropriate for systems that are flat and have no vertical
extent.  In a dimensionless system of units, the basis functions are
\begin{equation}
  \mu_{nm}(\mbox{\boldmath $R$})=\frac{2n+1}{2\pi}
  \left(\frac{1-\xi}{2}\right)^{3/2}P_{nm}(\xi)\exp(im\theta),
  \label{mubase}
\end{equation}
\begin{equation}
  \Phi_{nm}(\mbox{\boldmath $R$})=-\left(\frac{1-\xi}{2}\right)^{1/2}
   P_{nm}(\xi)\exp(im\theta),
  \label{phibase}
\end{equation}
where $\mbox{\boldmath $R$}=(R, \theta)$ is the position vector, $P_{nm}$
is the Legendre function, and $n$ and $m$ $(n\ge m)$ are the radial and
azimuthal ``quantum numbers", respectively.  In particular, positive values
of $m$ correspond to the number of arms in spiral patterns.  In equations
(\ref{mubase}) and (\ref{phibase}), the radial transformation
\begin{equation}
  \xi=\frac{R^2-1}{R^2+1}
\end{equation}
is used.  With these basis functions $(\mu_{nm}, \Phi_{nm})$, each pair of
which satisfies Poisson's equation, the density and potential of the system
can be expanded as
\begin{equation}
  \mu(\mbox{\boldmath $R$})=\sum_{nm} A_{nm}(t)\mu_{nm}(\mbox{\boldmath $R$}),
  \label{muexpand}
\end{equation}
\begin{equation}
  \Phi(\mbox{\boldmath $R$})=
  \sum_{nm} A_{nm}(t)\Phi_{nm}(\mbox{\boldmath $R$}).
  \label{phiexpand}
\end{equation}
The amplitude of the $(n, m)$-mode is calculated from the absolute
value of the expansion coefficients, $|A_{nm}(t)|$.  If a spatially
constant shape like a bar pattern emerges in a model disk, $A_{nm}(t)$
will be proportional to $\exp(-i\omega t)$, where $\omega$ is the complex
eigenfrequency, and $Im(\omega)$ will be almost zero in a nonlinear
regime.  Thus, the pattern speed for the $(n, m)$-mode is obtained from
$Re(\omega)/m$.  In practice, we pay attention to only the fastest growing
mode with $(n, m)=(2, 2)$.  The maximum number of radial expansion
coefficients, $n_{\rm max}$, is taken to be 16, and the number of
azimuthal expansion coefficients, $m_{\rm max}$, is set to be 2 with
only even values being used; that is, $m=0$ and 2.  Although we carried
out a simulation with $n_{\rm max}=32$, we found no difference between
the results with $n_{\rm max}=16$ and those with $n_{\rm max}=32$.  We
employ $N=100,000$ particles of equal mass.  The equations of motion
are integrated in Cartesian coordinates using a time-centered leapfrog
algorithm.  We employ a system of units such that $G=M=h=1$.  If these
units are scaled to physical values appropriate for the Milky Way, the
unit of time is $1.31 \times 10^7$ yr.

We first run a simulation until the bar has developed completely in
the disk, and then we continue the evolution, after growing a BH
according to equation (\ref{bhmass}).  Prior to adding a BH, we use a
timestep $\Delta t=0.1$ up to time $t=100$ when the bar is no longer
evolving.  After $t=100$, when we add a BH, we employ a timestep
$\Delta t=0.005$.  This choice of timestep was determined by
performing simulations with different values of $\Delta t$ and
requiring that the results of the integrations no longer depended on
$\Delta t$.  For the choice $\Delta t=0.005$, the total energy of the
system after the full growth of the BH was, in all cases,
conserved to better than four significant figures.

One might be concerned that our results will be affected by various
numerical approximations.  For example, our decision to use razor-thin
disks could enhance the influence of the black hole on the disk by
requiring the orbits of stars to remain in a single plane.  We
intend to examine this issue in future studies, but for now we note
that Norman \etal (1996) modeled disks both with and without
vertical extent, and did not find a significant difference in the
magnitudes of the central mass concentrations required for destroying
a bar.  On the other hand, by softening the black hole potential and
by forcing it to remain stationary at the origin, we may be
underestimating the response of the bar.  In reality, a black hole
near the center of a galaxy would generate a potential that is
essentially that of a point mass, and would ``wander'' about the
origin as it achieves equipartition with the background stars,
possibly enhancing the rate at which a bar would be destroyed (see,
\eg Quinlan \& Hernquist 1997).

\section{RESULTS}

To quantify the consequences of a black hole for a bar, we record
the amplitude of the azimuthal term in the density of the disk as
determined by the SCF expansion.  In Figure 1, we show the time
evolution of the bar amplitude for three values of the BH mass:
$M_{\rm BH}=0.01$, $0.005$, and $0.001$.  These choices bracket the
range in $M_{\rm BH}$ over which the black hole begins to have a
significant effect on the bar in our models.  In all cases, black hole
growth commenced at $t=100$, after the bar was fully developed, and
was completed by $t=110$.  The evolution beyond $t=110$ thus reflects
the influence of the black hole on the bar.

We can see from Figure 1 that the bar amplitude decreases with time
for $M_{\rm BH}=0.01$ and $0.005$ while it remains nearly constant to
the end of the simulation for $M_{\rm BH}=0.001$.  As a further
comparison, we also evolved a disk up to $t=300$ without growing a black
hole in it, and the outcome of this calculation is indicated by 
the uppermost curve at late times in Figure 1.  While not identical to
the case with a BH of mass $M_{\rm BH}=0.001$, the results of these two
experiments are sufficiently close that we cannot yet claim that a BH of
this low mass has a noticeable influence on the bar.

\begin{figure}
\figurenum{1}
\plotone{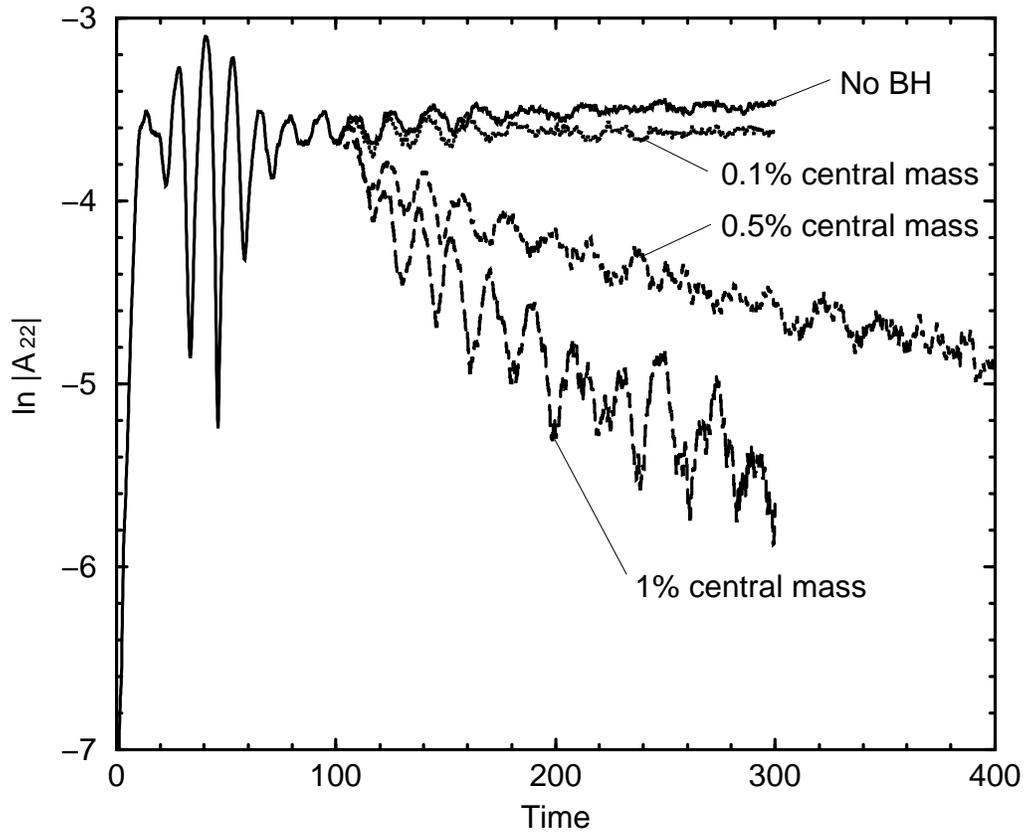}
\caption{Time evolution of the bar amplitudes of the fastest growing
mode, $|A_{22}|$, for $M_{\rm BH}=0.01$, $0.005$, and $0.001$.}
\end{figure}

Figure 1 demonstrates that the rate at which the bar is destroyed is
higher with increasing BH mass: the time required for bar dissolution
becomes shorter for increasingly more massive black holes.  In addition,
in cases where the influence of the black hole is significant, the bar
dissolves gradually with time.  In the experiments with black holes of
masses $M_{\rm BH}=0.01$ and $0.005$, the amplitude of the bar decays
nearly exponentially with time $\sim \exp (-t/\tau)$, once the BH is fully
developed.  From the decline of $\ln |A_{22}|$ with time, we estimate
decay times $\tau \sim 115$ and $\tau \sim 295$ for the simulations
with BH masses $M_{\rm BH}=0.01$ and $M_{\rm BH}=0.005$, respectively.
When scaled to values appropriate for the Milky Way, these correspond
to timescales $\sim 1.5 \times 10^9$ and $\sim 3.8 \times 10^9$
years, respectively.  Since these time intervals are small (but not
negligible) compared with the estimated ages of disk galaxies, we
tentatively conclude that black holes even with masses as small as
$0.5\%$ that of the disk can destroy a bar which formed at around
the same time as the disk.

\begin{figure}
\figurenum{2}
\plotone{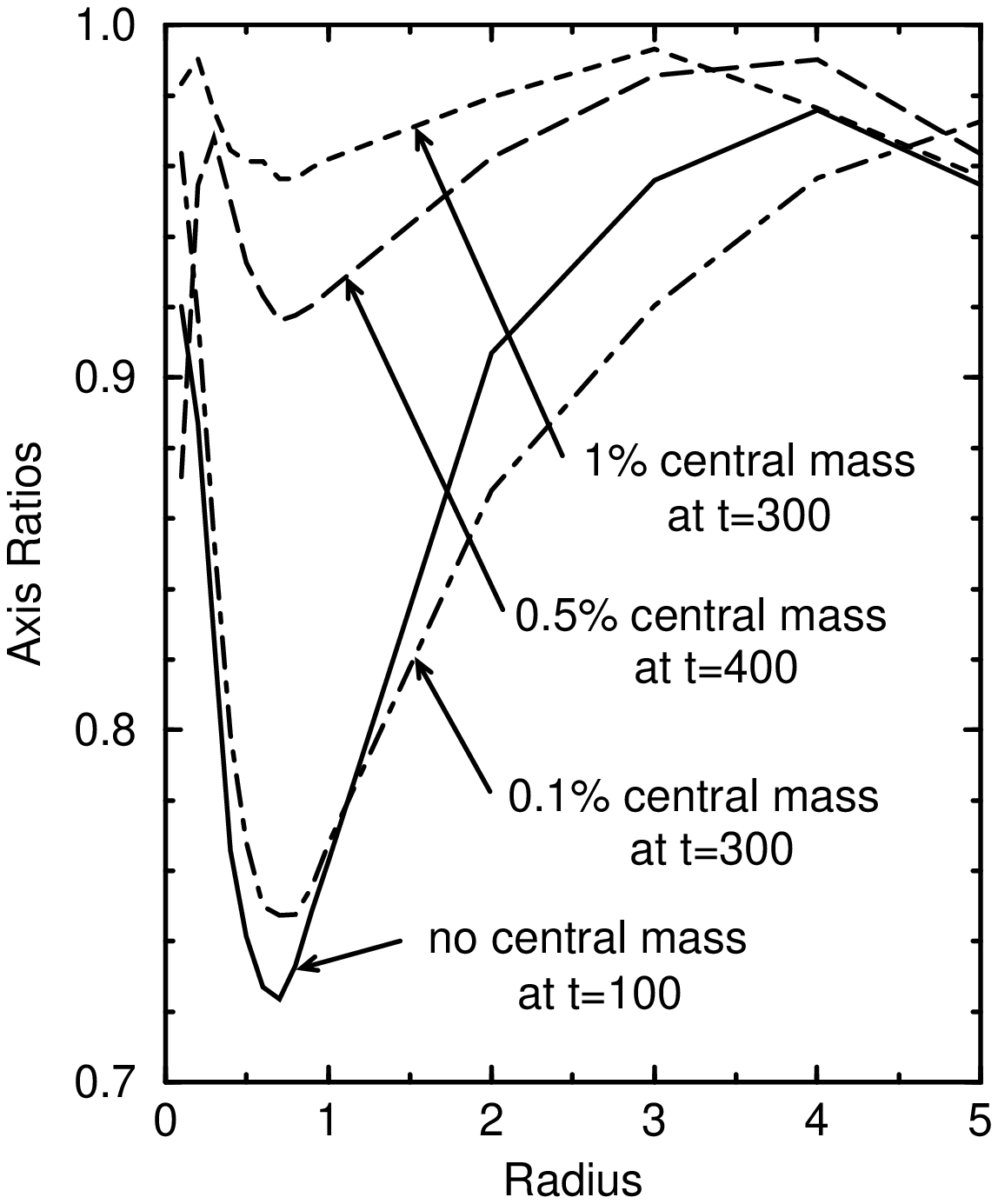}
\caption{Change in axis ratios from $t=100$ to the end of the simulations
for $M_{\rm BH}=0.01$, $0.005$, and $0.001$.}
\end{figure}

These arguments are supported by examining the structural properties of 
the bars in each of our simulations following the growth of the black
hole.  We determine the axis ratios of the bars by calculating the
moment of inertia tensor for particles included in a specified radius,
and use this information to derive the axis ratio at that radius.  In
Figure 2, we show the axis ratio of the bars thus computed for the
four experiments in Figure 1.  As is apparent from Figure 2, the axis
ratio has become $\gtrsim$ 0.96 for $M_{\rm BH}=0.01M$ at $t=300$ and
$\gtrsim$ 0.92 for $M_{\rm BH}=0.005M$ at $t=400$ from $\sim 0.72$ at
$t=100$ within the bar regions.  On the other hand, the axis ratio has
changed little from $t=100$ to $t=300$ for $M_{\rm BH}=0.001M$.  Thus,
the bar can be destroyed within a short time scale as compared to a
Hubble time if the BH mass exceeds about 0.5\% of the disk mass.

\section{DISCUSSION AND CONCLUSIONS}

We have shown that a massive central BH can dissolve a bar within a
short time scale if the BH is as massive as about 0.5\% of the disk
mass.  This means that the minimum BH mass necessary for bar dissolution
would be of order $10^{8.5} M_\odot$ for a typical disk galaxy.  This
minimum BH mass is an order of magnitude smaller than that implied from
the results of Norman \etal (1996), if we associate the central mass
concentrations in their models with black holes.  Since our minimum BH
mass is not extremely large compared with the BH masses suggested by
observations, bar dissolution should not be a rare event but could occur
at some unexceptional rate in real barred galaxies.

It is not yet clear why we obtain a minimum mass for bar destruction
that is so smaller than that which is suggested by the Norman \etal
results.  Our simulations differ from theirs in several respects,
and we do not know which difference is most responsible for our lower
value of this BH mass.  We suspect that a likely culprit is the
difference in the galaxy models employed in the two studies.  Norman
\etal employ multi-component models in which the disk is represented
by a Kuzmin-Toomre mass profile (Kuzmin 1956; Toomre 1963) and $25\%$
of the mass resides in concentric bulge and ``core'' components that
are modeled as Plummer spheres.  To mimic the effects of gas inflow,
the scale-length of the Plummer sphere representing the central mass
concentration is reduced slowly with time.  In cases of interest, the
core component contains up to $10\%$ of the system mass.

In our simulations, the disks have exponential profiles, and we allow
the black hole mass to grow slowly with time, but we do not alter the
scale-length of the black hole potential.  While it is conceivable
that the important difference is the manner in which the density of
the central mass concentration is varied (Norman \etal fix the mass
while we fix the scale-length) a more likely possibility is our use of
a disk profile that is significantly more concentrated than that in
the Norman \etal simulations.  For example, it is possible that a
relatively larger fraction of the stars supporting the bars in our
simulations pass sufficiently near that central mass that they could
be strongly perturbed by it.  Indeed, Norman \etal argue that the
critical mass will likely depend sensitively on the properties of the
galaxy and the bar.  However, there are other technical differences
between the two sets of calculations and we have not explored
parameter space in sufficient detail to show that the structure of the
galaxy is primarily responsible for our smaller critical BH mass.

For cold systems in which rotation is dominant, the introduction of a
softening length can alter the dynamics of a disk.  Earn \& Sellwood
(1995) have demonstrated that for an isochrone disk, the growth rate
of the fastest growing two-armed mode obtained with their smallest
softening length is still 20\% smaller than that derived from linear
analysis.  On the other hand, if an SCF method is used, the growth
rate is in excellent agreement with that predicted by linear theory.
As a result, the estimated BH mass may differ from that which is
actually needed to destroy a bar, when a numerical code requiring
force softening is used.

Another interesting difference between our results and those of Norman
\etal (1996) is the rate at which bars are destroyed in response to
the central mass concentrations.  In our models, as indicated by
Figure 1, the amplitude of the bar declines smoothly and slowly with
time.  Norman \etal find that the bars in their simulations are
destroyed relatively abruptly when the central mass exceeds some
critical value, and they argue that this results from chaotic behavior
caused by the modification to the potential by the central mass.  It
is unclear if this difference is driven by the different mass models
employed for the galaxies in the two studies, or if it is numerical in
origin.  However, this difference may be related to the physical
process by which the bar is destroyed.  Several authors have argued
that a central black hole can scatter stars on orbits supporting a
bar, and that a bar would be gradually eroded by this process (\eg
Norman 1984; Norman, May, \& van Albada 1985; Gerhard \& Binney
1985).  In principle, this interpretation can account for the evolution
seen in our models.  If this is indeed the case, even a relatively small
BH mass could affect the structure of a bar, if a sufficient number of
orbits pass within the black hole's ``sphere of influence.''
In our simulations, a BH with $M_{\rm BH}=0.001$ was unable to destroy
a bar.  It is possible that this outcome was unduly influenced by
poor resolution near the center of our disks and our suppression of
black hole ``wandering''.  Clearly, simulations with a larger number
of particles and a greater degree of physical realism will be required
to work out the true nature of the mechanism of bar dissolution.

\acknowledgments  This work was supported in part by a Grant-in-Aid for
Scientific Research from the Ministry of Education, Science, Sports, and
Culture of Japan (09740172), and by NASA theory grant NAG5-3059 and the
NSF under grant ASC 93-18185.

\end{document}